\documentclass{article}
\usepackage{spconf,amsmath,graphicx}
\usepackage{siunitx, booktabs, tabularx, multirow, subcaption}
\usepackage{fancyhdr}

\title{Measuring Soft Biometric Leakage in Speaker De-Identification Systems}

\name{Seungmin Seo, Oleg Aulov, P. Jonathon Phillips \thanks{This research is based upon work supported by the Intelligence Advanced Research Projects Activity (IARPA) \emph{Anonymous Real-Time Speech (ARTS)} program.}}
\address{National Institute of Standards and Technology, Gaithersburg, MD, USA}

\fancypagestyle{firstpage}{
  \fancyhf{} 
  \fancyfoot[C]{\small This work has been submitted to the IEEE for possible publication. 
  Copyright may be transferred without notice, after which this version may no longer be accessible.}
}

\begin{document}
%
\thispagestyle{firstpage}
\maketitle
\begin{abstract}

We use the term re-identification to refer to the process of recovering the original speaker’s identity from anonymized speech outputs. Speaker de-identification systems aim to reduce the risk of re-identification, but most evaluations focus only on individual-level measures and overlook broader risks from soft biometric leakage. We introduce the Soft Biometric Leakage Score (SBLS), a unified method that quantifies resistance to zero-shot inference attacks on non-unique traits such as channel type, age range, dialect, sex of the speaker, or speaking style. SBLS integrates three elements: direct attribute inference using pre-trained classifiers, linkage detection via mutual information analysis, and subgroup robustness across intersecting attributes. Applying SBLS with publicly available classifiers, we show that all five evaluated de-identification systems exhibit significant vulnerabilities. Our results indicate that adversaries using only pre-trained models—without access to original speech or system details—can still reliably recover soft biometric information from anonymized output, exposing fundamental weaknesses that standard distributional metrics fail to capture.

\end{abstract}
\begin{keywords}
speaker de-identification, soft biometrics, re-identification, speaker biometric leakage
\end{keywords}
\section{Introduction}
\label{sec:intro}

In addition to linguistic content, speech signals encode speaker characteristics such as identity, age range, sex, and dialect \cite{FairSpeech2024}. Speaker de-identification (SDID) systems aim to transform speech to remain intelligible while masking these traits \cite{VoicePrivacy2020,Tomashenko2022,Li2025HLTCOE}.

Current SDID evaluations focus primarily on individual-level metrics such as equal error rate (EER) for speaker verification, measuring how well single utterances can be linked to source speakers \cite{Li2025SpecWavAttack, Li2025HLTCOE, Mawalim2025TitaNetLarge, Zhang2025AugmentedFeatureAttack}. However, recent advances in self-supervised learning have enabled highly accurate inference of demographic attributes from speech using publicly available classifiers \cite{yang2025demographic}. This fundamentally alters the de-identification landscape: attackers can now extract sensitive biometric information without accessing original recordings or understanding system internals.

Traditional metrics overlook these risks. EER measures speaker linkage but ignores soft biometric inference. Distributional measures such as KL divergence can also mislead—identical marginal distributions may coexist with high soft biometric attribute predictability, while shifts in distributions might preserve systematic patterns that are exploitable. As a result, current evaluations leave important questions unanswered: \textbf{Can attackers still infer soft biometrics from anonymized speech?} \textbf{Do these systems create systematic artifacts that can be exploited}? 

We introduce the \textbf{Soft Biometric Leakage Score (SBLS)}, a unified method that measures exposure through zero-shot attack scenarios. SBLS integrates three components: (i) zero-shot attribute inference quantifying direct soft biometric prediction, (ii) systematic linkage detection capturing exploitable transformation patterns via mutual information, and (iii) subgroup robustness identifying differential exposure across attribute intersections. Evaluation of five SDID systems reveals measurable leakage that standard metrics fail to capture, demonstrating the need for more comprehensive assessment methods in SDID research.

\section{Soft Biometric Leakage Score}
\label{sec:sbls}

We define a SBLS as a combination of three interpretable components:
\begin{align}
\mathrm{SBLS} &= \alpha P_{\text{attr}} + \beta P_{\text{assoc}} + \gamma P_{\text{subgroup}}, \nonumber\\ 
\quad & \alpha,\beta,\gamma \ge 0,\quad \alpha+\beta+\gamma = 1. 
\end{align}

\noindent All components are in $[0,1]$ and are computed using a fixed zero-shot attacker. We align predictions to true labels to avoid permutation artifacts and capture both average-case and worst-case leakage. The weights $\alpha,\beta,\gamma$ can be fixed (e.g., equal weighting), tuned on a development set, or chosen to emphasize particular components. In this work, we set $\alpha=\beta=0.4$ and $\gamma=0.2$ to slightly downweight subgroup robustness, though the choice is heuristic.

\subsection{Zero-Shot Attribute Inference}
\label{subsec:attack}

Let $A$ be the set of attributes (e.g., $A=\{\text{Male/Female label},\allowbreak \text{age group}\}$). For each attribute $a\in A$ with $K_a$ classes, the frozen attacker outputs class scores $S_a \in \mathrm{R}^{N\times K_a}$ on the anonymized dataset. We compute one-vs-rest AUC for each class $k$ as $\mathrm{AUC}_{a,k}$ from scores $S_a[:,k]$. 

To ensure permutation invariance, we align predicted score columns to true labels via the optimal assignment:
\begin{equation}
    \mathrm{mAUC}_a^\star \;=\; \frac{1}{K_a}\,\max_{\pi \in S_{K_a}}\;\sum_{k=1}^{K_a}\mathrm{AUC}_{a,\pi(k)},
\end{equation}
where $S_{K_a}$ denotes all permutations of $K_a$ elements. The chance-level baseline for one-vs-rest AUC is $0.5$. We convert to leakage score by normalizing and inverting:
\begin{equation}
    P_{\text{attr}} = 1 - \frac{1}{|A|}\sum_{a\in A} \frac{\max\{0,\;\mathrm{mAUC}_a^\star - 0.5\}}{0.5}
    \;\in[0,1]
\end{equation}
Here $P_{\text{attr}}{=}1$ indicates chance-level zero-shot inference (low leakage) and $P_{\text{attr}}{=}0$ indicates near-perfect attribute recoverability (high leakage). When only hard predictions are available, we substitute macro balanced accuracy for $\mathrm{mAUC}_a^\star$ and normalize by the chance-level baseline $1/K_a$.

\subsection{Systematic Linkage Detection}
\label{subsec:systematic}

Let $\hat A_a^\star$ denote hard predictions obtained via $\arg\max$ over the permutation-aligned scores from the previous step.

We compute normalized mutual information to measure systematic dependence:
\begin{equation}
\begin{aligned}
    \tilde I_a \;=\; \frac{I\!\bigl(A_a;\,\hat A_a^\star\bigr)}{\log K_a}\;\in[0,1],\\
    P_{\text{assoc}} \;=\; 1 \;-\; \frac{1}{|A|}\sum_{a\in A} \tilde I_a,
\end{aligned}
\end{equation}
where $I(A_a; \hat A_a^\star)$ is the mutual information between true and predicted attributes. In practice, we estimate $\tilde I_a$ from the confusion matrix between $(A_a, \hat A_a^\star)$ using standard entropy calculations. High $\tilde I_a$ indicates strong systematic dependence (higher leakage), while $\tilde I_a \approx 0$ suggests that predictions contain little information about true attributes beyond random chance.

\subsection{Subgroup Protection and Consistency}
\label{subsec:subgroup}

SDID systems may provide strong average protection while systematically failing for specific soft biometric attribute subgroups. Let $G$ denote the set of soft biometric attribute intersections (e.g., \{young male, adult female, senior male, ...\}). For each subgroup $g \in G$ with sufficient samples (we require $n_g \geq 10$), we evaluate zero-shot attribute inference performance within that subgroup. We define the subgroup-specific leakage as:
\begin{equation}
    L_g \;=\; \frac{1}{|A|}\sum_{a\in A} \frac{\max\{0,\;\mathrm{mAUC}_{a,g}^\star - 0.5\}}{0.5},
\end{equation}
where $\mathrm{mAUC}_{a,g}^\star$ is computed using only samples from subgroup $g$. The overall subgroup protection score combines both worst-case performance and consistency considerations:

\begin{equation}
    P_{\text{subgroup}} = \omega \bigl(1 - \max_{g \in G} L_g\bigr) + (1-\omega) \frac{\min_{g \in G}(1 - L_g)}{\max_{g \in G}(1 - L_g)},
\end{equation}
where $\omega \in [0,1]$ balances worst-case protection against consistency across groups . We set $\omega = 0.7$ to emphasize worst-case performance while still penalizing large disparities. The first term equals $0$ when any subgroup has perfect leakage, while the second term approaches $0$ when protection varies dramatically across subgroups.

\begin{table*}[t]
\centering
\caption{Summary of SDID systems, methods, and training datasets.}
\label{tab:full_system_summary}
\resizebox{0.9\linewidth}{!}{%
\small
\renewcommand{\arraystretch}{1.15}
\setlength{\tabcolsep}{5pt}
\begin{tabular}{p{2cm} p{4cm} p{5.3cm} p{5cm}}
\toprule
\textbf{System} & \textbf{Method} & \textbf{Key Components} & \textbf{Training Data} \\
\midrule
VOXLET \newline(Galois)& Latent perturbation + vocoder & wav2vec 2.0 encoder, HiFiGAN 2.0 vocoder, DP noise in latent space & Encoder: LibriSpeech, LibriVox, TIMIT; Vocoder: DAPS + MIT IRs + REVERB/ACE \\
\midrule
RASP \newline(Honeywell)& Disentangled autoencoder & HuBERT (content); speaker replaced with pseudo-embedding (ControlVCSpawn); pitch/energy via variance encoder & HuBERT: LibriSpeech; Speaker encoder: VoxCeleb1 + LibriSpeech; Full model: VCTK \\
\midrule
SHADOW \newline(JHU)& Token conversion + speaker swap & EnCodec + Wav2Vec2 tokens; PLDA for speaker swapping; FreeVC module & Main: Libri-Light, MLS, LibriTTS-R; FreeVC: VCTK; PLDA: VoxCeleb2, LibriTTS-R \\
\midrule
PHORTRESS \newline(SRI)& Articulatory synthesis SPARC & Source extractor, EMA inverter, speaker encoder, DDSP vocoder, fabricated speaker embedding & GigaSpeech, LibriTTS-R, VCTK; Speaker embedding: Pyannote \\
\midrule
kNN-VC \newline(LLNL-baseline)& $k$-NN in WavLM space & Frame-wise regression using nearest neighbors; re-synthesized via HiFiGAN; fusion across pseudo-profiles & Latent space: WavLM; Vocoder: HiFiGAN trained on LibriSpeech \\
\bottomrule
\end{tabular}
}
\end{table*}

\section{Experimental Setup}
\label{sec:setup}

\subsection{Dataset and Demographics}

We evaluate on the Mixer 3 corpus subset, selected for its rich demographic annotations and lengthy conversational speech characteristics that challenge SDID systems. We retained only native American English speakers with at least five recording sessions, yielding 223 speakers, of whom 180 of these have ten or more sessions. In total, 2,983 speech segments were generated with at least 10, 30, and 60 seconds of detected speech activity. Demographic statistics are presented in Table~\ref{tab:speaker_agegroup}.

\begin{table}[th]
    \centering
    \caption{Number of unique speakers by age group and Male/Female designation.}
    \vspace*{0.5\baselineskip}
    \label{tab:speaker_agegroup}
    \resizebox{0.5\linewidth}{!}{%
        \begin{tabular}{ccc}
        Age Group & Male & Female \\
        \hline
        Young (17--24) & 17 & 15 \\
        \hline
        Adult (25--54) & 57 & 104 \\
        \hline
        Senior (55--85) & 7 & 23 \\
        \hline
        \end{tabular}
    }
\end{table}

\subsection{Speaker De-Identification Systems}

The five SDID systems in this study were submitted to NIST for evaluation—all developed under the IARPA ARTS program\footnote{www.iarpa.gov/research-programs/arts}—including four performer submissions and one baseline system built by a Test \& Evaluation partner. See Table \ref{tab:full_system_summary} for details. Note that no system descriptions are publicly available at the time of this writing, so the references reflect relevant work by the same researchers. \cite{wang2025diffattack, wang2024toward, sang2025unipet,xinyuan2024hltcoe}

\subsection{SBLS Component Implementation}

\textbf{Zero-Shot Attribute Leakage:} We compute one-vs-rest AUC for each soft biometric attribute using the classifier's soft prediction scores. After optimal permutation alignment, we normalize by the chance-level baseline (0.5 for AUC) and average across attributes to obtain $P_{\text{attr}}$.

\noindent\textbf{Systematic Linkage Detection:} We calculate normalized mutual information between true and predicted soft biometric labels using the confusion matrix approach for robust estimation. The mutual information is normalized by $\log K$ where $K$ is the number of classes to ensure bounded scores.

\noindent\textbf{Subgroup Protection:} For each soft biometric intersection, we compute within-group zero-shot attribute inference performance. The component combines worst-case protection (maximum leakage across subgroups) with consistency analysis (protection ratio) using $\omega = 0.7$ to emphasize worst-case performance.

\section{Evaluation Results}
\label{sec:results}

\subsection{SBLS Rankings and System Comparison}

Table~\ref{tab:SBLS_scores} presents the comprehensive Soft Biometric Leakage Score (SBLS) evaluation across all SDID systems. PHORTRESS achieves the highest overall score (0.856), driven by its exceptional zero-shot attribute leakage ($P_{\text{attr}}$ = 0.994) that approaches chance-level performance for demographic classifiers. However, its subgroup protection remains moderate (0.531), indicating systematic vulnerabilities persist for specific demographic intersections. VOXLET exhibits the poorest leakage protection (SBLS = 0.661), with substantial zero-shot soft biometric attribute leakage ($P_{\text{attr}}$ = 0.690) and severe subgroup vulnerabilities ($P_{\text{subgroup}}$ = 0.332).

\begin{table}[t]
    \centering
    \caption{Soft Biometric Leakage Score (SBLS) and components for evaluated SDID systems. Higher values indicate better leakage protection, with ($\alpha=0.4, \beta=0.4, \gamma=0.2$).}
    \label{tab:SBLS_scores}
    \resizebox{0.7\linewidth}{!}{%
        \begin{tabular}{lcccc}
        \toprule
        SDID System & $P_{\text{attr}}$ & $P_{\text{assoc}}$ & $P_{\text{subgroup}}$ & \textbf{SBLS} \\
        \midrule
        PHORTRESS   & 0.994 & 0.998 & 0.531 & 0.903 \\
        SHADOW      & 0.936 & 1.000 & 0.501 & 0.874 \\
        kNN-VC      & 0.877 & 0.993 & 0.604 & 0.869 \\
        RASP        & 0.910 & 0.995 & 0.435 & 0.849 \\
        VOXLET      & 0.690 & 0.950 & 0.332 & 0.723 \\
        \bottomrule
        \end{tabular}
    }
    \vspace{-0.5em}
\end{table}

Notably, all systems achieve high systematic linkage detection scores ($P_{\text{assoc}} \geq 0.950$), indicating successful avoidance of exploitable transformation patterns. This suggests that while soft biometrics related to demographic information may leak through direct inference, attackers cannot easily learn systematic rules to reverse-engineer the de-identification process.

\subsection{Zero-Shot Soft Biometric Attribute Inference Vulnerabilities}

Table~\ref{tab:demographic_inference} reports the performance of off-the-shelf classifiers applied to anonymized speech. The AUC values, which range only modestly above chance in most cases, suggest that zero-shot soft biometric attribute inference is possible to a limited extent, though the strength of these signals varies across both attributes and systems.

\begin{table}[t]
    \centering
    \caption{Zero-shot soft biometric attribute inference performance using VoxProfile classifier. AUC values represent classifier accuracy on anonymized speech for Male/Female binary attribute, and for age group (0.5 = chance level, 1.0 = perfect inference).}
    \label{tab:demographic_inference}
    \resizebox{\linewidth}{!}{%
        \begin{tabular}{lcccc}
        \toprule
        SDID System & M/F AUC & Age AUC & M/F Vulnerability & Age Vulnerability \\
        \midrule
        PHORTRESS   & 0.501 & 0.505 & Minimal & Minimal \\
        RASP        & 0.513 & 0.577 & Low & Moderate \\
        SHADOW      & 0.538 & 0.526 & Low & Low \\
        kNN-VC      & 0.543 & 0.580 & Moderate & Moderate \\
        VOXLET      & 0.721 & 0.589 & Noticeable & Moderate \\
        \bottomrule
        \end{tabular}
    }
\end{table}

Male/Female attribute-related leakage differs across systems. While VOXLET shows the highest leakage (AUC = 0.721), indicating that some acoustic correlates of this attribute (e.g., pitch and formant cues) are preserved, other systems remain much closer to chance. For example, PHORTRESS achieves near-chance performance (AUC = 0.501), effectively reducing the inference risk.

Age-related leakage appears more uniformly protected, with AUC values between 0.505 and 0.589 across systems. These results suggest that current anonymization approaches generally obscure age-related acoustic features, though some residual vulnerability remains, particularly for classifiers with moderate performance.

\begin{figure}[t]
    \centering
    \includegraphics[width=\linewidth]{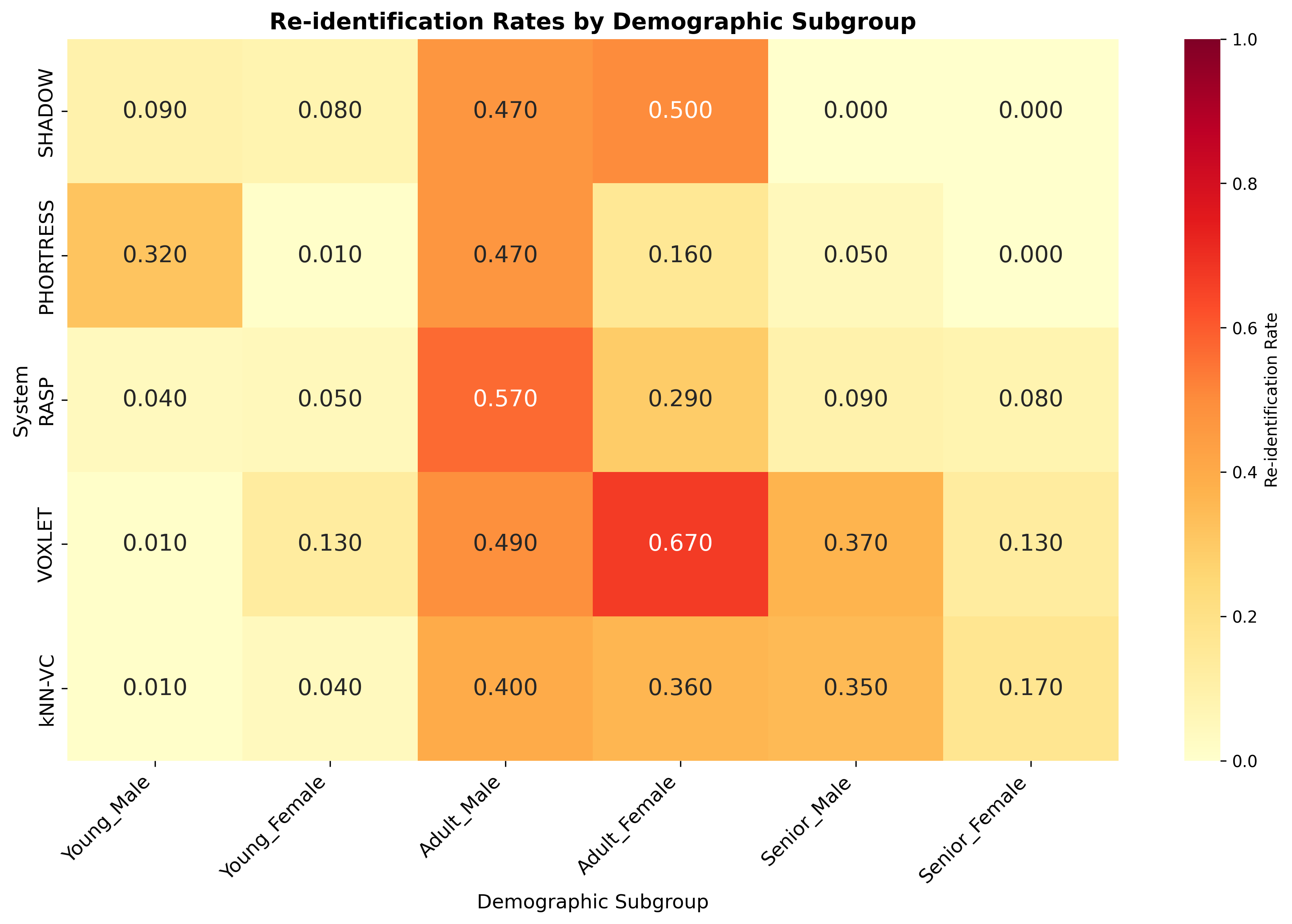}
    \caption{Heatmap of re-identification rates across soft biometric attribute subgroups and SDID systems. Darker colors indicate higher vulnerability. Adult speakers show consistent re-identification risk across systems.}
    \label{fig:vulnerability_heatmap}
\end{figure}

\begin{figure}[t]
    \centering
    \includegraphics[trim=150 20 250 20,clip, width=\linewidth]{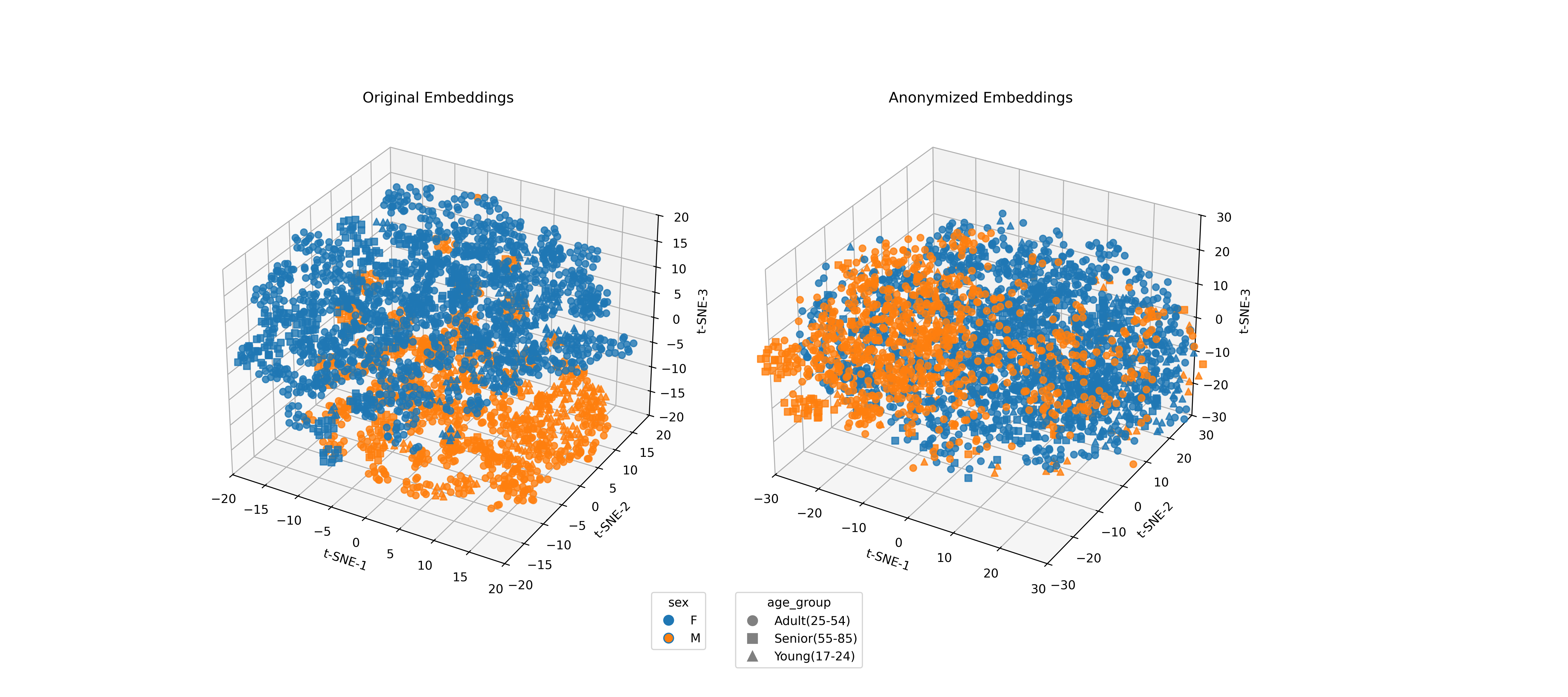}
    \caption{t-SNE visualization (VOXLET): (a) Original embeddings show clear soft biometric attribute clustering, (b) Anonymized embeddings maintain visible structure enabling distribution recovery. Points colored by the original speaker's Male/Female label, shaped by the original age group.}
    \label{fig:tsne}
\end{figure}

\subsection{Systematic Linkage and Pattern Avoidance}

All systems demonstrate strong resistance to systematic exploitation ($P_{\text{assoc}} \geq 0.950$). This indicates that anonymization errors appear random rather than following predictable transformation rules. 

High $P_{\text{assoc}}$ scores across all systems suggest that:
\begin{itemize}
    \item Attackers cannot learn exploitable mapping functions (e.g., "system always transforms male voices to female predictions")
    \item Soft biometric attribute inference failures occur through direct classifier application rather than systematic pattern exploitation
    \item Anonymization systems successfully avoid creating deterministic soft biometric attribute transformations
\end{itemize}

\noindent This finding has important implications for adversarial robustness: while zero-shot classifiers pose immediate threats, more sophisticated attacks that attempt to learn system-specific transformation patterns are unlikely to succeed.

\subsection{Subgroup Protection Analysis}

Figure~\ref{fig:vulnerability_heatmap} shows that the adult subgroup exhibits the highest re-identification rates across all systems, ranging from 39.8\% to 66.9\%. This pattern suggests that the acoustic characteristics of middle-aged voices may be more distinctive and therefore harder to anonymize effectively. The effect may also be influenced by the dataset distribution, since adults represent the largest portion of speakers.

\subsection{Embedding Space Analysis}

To investigate the mechanism of soft biometric attribute leakage, we examine the structure of anonymized embedding spaces using t-SNE visualization. Figure~\ref{fig:tsne} shows that attribute-related structure persists: while individual speakers are not identifiable, the embeddings remain organized along soft biometric attributes. This residual organization provides the basis for the distribution-recovery attacks observed in our evaluation.

\section{Conclusion}

The widespread availability of high-accuracy classifiers for soft biometric attributes makes identity leakage in speech anonymization a measurable concern, as adversaries can recover such information without specialized knowledge. Our results, though limited to a single corpus and two attributes, show that current de-identification systems do not fully mitigate soft biometric leakage. Metrics such as Soft Biometric Leakage Score (SBLS) provide a practical means to quantify exposure under realistic threat models and to assess the robustness of anonymization methods.

\section{Acknowledgments and Disclaimers}
\label{sec:ack}

This research is based upon work supported by the Office of the Director of National Intelligence (ODNI), Intelligence Advanced Research Projects Activity (IARPA), Anonymous Real-Time Speech (ARTS) research program, under Interagency Agreement (IAA) with NIST IARPA-20001-D250300042. The views and conclusions contained herein are those of the authors and should not be interpreted as necessarily representing the official policies or endorsements, either expressed or implied, of the ODNI, IARPA, NIST or the U.S. Government.

Certain equipment, instruments, software, or materials are identified in this paper in order to specify the experimental procedure adequately. Such identification is not intended to imply recommendation or endorsement of any product or service by NIST, nor is it intended to imply that the materials or equipment identified are necessarily the best available for the purpose. These opinions, recommendations, findings, and conclusions do not necessarily reflect the views or policies of NIST or the United States Government.

\bibliographystyle{IEEEbib}
\bibliography{manuscript}

\end{document}